\documentclass[epj,onecolumn]{svjour}
\usepackage{graphics}
\usepackage{lipsum}
\usepackage{mathtools}
\usepackage{amsmath}
\usepackage{cuted}
\begin{document}
\title{{Cylindrically symmetric and plane-symmetric solutions in $f(R)$ theory via Noether symmetries}}
\author{Işıl Başaran Öz\inst{1}\thanks{\emph{e-mail:} ibasaranoz@gmail.com (corresponding author)}\and Kazuharu Bamba\inst{2}\thanks{\emph{e-mail:}  bamba@sss.fukushima-u.ac.jp}}
\institute{Bayındır Neighborhood 324 Street No:12, Muratpaşa, 07030 Antalya, Turkey \and Division of Human Support System, Faculty of Symbiotic Systems Science, Fukushima University, Fukushima 960-1296, Japan}
\date{Received: date / Revised version: date}
\abstract{
The $f(R)$ theory is considered for static cylindrically symmetric and plane-symmetric spacetimes. In order to find solutions to the field equations of these models, the Noether symmetry method is used. First, we examine the GR case for cylindrically symmetrical space-time with the $w=-1$ dark energy state. Then, with the assumption of $f(R) = f_0R^n$, cases with matter and non-matter are examined and general solutions are determined for both space-times. Thus, it is shown that inclusive new solutions are obtained, considering the Noether symmetric conditions. In addition, the GR limit for each cases are examined.
} 
\maketitle
\section{Introduction}
\label{intro}

At the end of the last century, the discovery that our universe is accelerating the expansion as a result of supernova Type Ia observations made by two independent groups has been an important discovery in cosmology \cite{Perlmutter1998,Riess1998}. It is thought that there must be some kind of energy with negative pressure as the source of this expansion. Thus, the $\Lambda CDM$ model has been widely accepted as the standard model of cosmology. However, this model suffers from fine-tuning and coincidence problems, and in addition, the nature of dark energy cannot be fully explained \cite{Weinberg1972,padmanabhan2007,Copeland2006,Durrer2008}. For this reason, modified gravity theories have been extensively studied in the literature in order to create a more general theory of gravity that can solve the problems of the standard cosmological model and enable us to better understand the structure of the universe \cite{Clifton2012,Bamba2012,Capozziello2011,Aydiner2022}.

One of the most fundamental modifications is that based on changing the geometry of the GR, where the Ricci scalar R is replaced by the f(R) function \cite{Copeland2006,Nojiri2010,Nojiri2017,Sotiriou2008,DeFelice2010}. The $f(R)$ theory has applications to explain the cosmic acceleration in the early universe, for example, in Ref. \cite{Starobinsky1980,Carroll2003,Nojiri2003,Capozziello2003,Capozziello2010}. The thermodynamics of the visible horizon \cite{Bamba2009,Bamba2010a} and the curvature singularity that occurs during the stellar collapse process are also studied in $f(R)$ gravity \cite{Bamba2011}. For the $f(R)$ gravity, future transitions of the phantom divide line for dark energy and the variation of acceleration in the expansion are shown in Ref. \cite{Bamba2010b}. Also, reconstruction of the f(R) gravity models with bounce cosmology is performed in \cite{Bamba2013}.

The anisotropies in the CMB observations suggest that the universe is homogeneous but anisotropic \cite{Hu2002}. For this reason, it is important to examine the cylindrically symmetric and plane-symmetric space-times within the framework of both GR and modified gravitational theories \cite{Trendafilova2011,Sharif2012b,Houndjo2012,FarasatShamir2014,Banerjee2018,Setare2009,Ali2019,Nurbaki2020,Malik2022}.  A recent comprehensive review has examined cylindrical systems in GR \cite{Bronnikov2019}. Once again, these backgrounds are very useful for studying compact objects and investigating gravitational waves \cite{Ali2019,Sharif2014,Bhatti2017,Yousaf2020,Lemos2008,Sharif2010}. Also, since cylindrical symmetry has an important geometric structure, it is often used for string theory and wormhole studies \cite{Momeni2017,Azadi2008,Bronnikov2019}. Therefore, it is valuable to obtain analytical solutions for cylindrical and plane-symmetric space-times in GR and/or modified gravitational theories.

Modified theories are highly non linear, so symmetry method is considered to be a powerful tool for calculating the solutions of their field equations. An imperative feature of Lie theory is Noether symmetry approach. Emmy Noether in 1918 \cite{Noether1918} determined that there is a conserved quantity for every corresponding symmetry. The importance of conserved quantity is because of laws of conservation, such as energy, momentum, angular momentum, etc., that give the conserved quantity for a given dynamical system.
In the literature, there are $f(R)$ and some modified gravitational theories solutions studied by the Noether symmetry approach for spherical symmetric space-time \cite{Capozziello2007,Capozziello2012,Bahamonde2018,Tsamparlis2018,Paliathanasis2011,Dimakis2017,Camci2016}. Some black hole solutions investigated under $f(R)$ theory using Noether symmetry are also available in the literature \cite{Camci2020,Darabi2013}. In our previous work, analytical solutions were obtained for some unfamiliar EoS parameters in the GR case for cylindrically symmetric space-time \cite{Oz2021}. In this work, we wanted to investigate the Noether symmetric general solutions for f(R) gravity with a cylindrical and plane symmetrical background. We concentrated on obtaining more inclusive general solutions for a model of type $f(R)=R^n$ rather than doing examined under various assumptions.

The article is organized as follows. In the first section, we construct the theory of $f(R)$ gravity with a static cylindrical symmetric and a static cylindrical symmetric background. Then, we obtained the solutions respectively in the following subsections. Finally, in the conclusion section, we summarize the findings of this study. Additionally in the appendix, first we explain the Noether symmetry approach, then give the Noether symmetry equations and vector components respectively for each cases.
\section{Field Equations and solutions}
\label{sec:FEq}

In this section, we consider the construction of cosmological models within the framework of the metric f(R) theory. The $f(R)$ theory of gravity is an interesting and relatively simple alternative to the modification of the GR. There are some basic works on the metric $f(R)$ gravity see here \cite{Sotiriou2008,Azadi2008,Sawicki2007,Sharif2012a}, and for alternative theories of gravity, including a comprehensive analysis of f(R) gravity, see also \cite{Clifton2012,Nojiri2010,Nojiri2017,DeFelice2010}. Starting from the action GR and containing the term $f(R)$ instead of the R term and the matter term Sm, the 4-dimensional total action for the $f(R)$ theory of gravity takes the following form:

\begin{equation}\label{actf(R)}
\mathcal{S}=\int{d^4x \sqrt{-g}\left[\frac{1}{2 \kappa}f(R)+\mathcal{L}_m\right]},
\end{equation}
where $\kappa=8\pi G$ and $\mathcal{L}_m$ is Lagrangian related to matter content of any kind in the universe. $f(R)$ is a function dependent on the scalar curvature $R$, while $ f(R)=R $ the theory is clearly reduced to the General Relativity \cite{Sotiriou2008}.

\subsection{Static cylindrically symmetric space-time} \label{ssec:FEq1}
The static cylindrical symmetric space-time metric is as follows \cite{Momeni2009,Shamir2014}.
\begin{equation}\label{sss-metric}
  ds^2=A(r)dt^2-dr^2-B(r)( d\theta^2+\alpha^2 dz^2),
\end{equation}
where $ A(r) $ and $ B(r) $ are the metric coefficients of taken depending on the radial coordinate  $ r $.
For such higher-order modified theories, with the help of the Lagrangian Multiplier Method and the integration by parts method, which are used to obtain the Lagrangian, the point-like Lagrangian is derived as follows \cite{Capozziello1994}.
\begin{eqnarray}\label{lag}
& &\mathcal{ L}=-\frac{\alpha f_{R}}{\sqrt{A}}A'B'-\alpha f_{RR}\frac{B}{\sqrt{A}}A'R'-\frac{\alpha f_{R}}{2B}\sqrt{A}B'^2\nonumber \\
& & \qquad    -2\alpha f_{RR}\sqrt{A}B'R'+\alpha \sqrt{A}B[f-R f_{R} -\kappa \rho_{0} A^{-\frac{1+w}{2w}}], \nonumber \\
& & \qquad (w \neq 0)
\end{eqnarray}
where ($'$)denotes the derivative with respect to $r$ and $f_R\equiv df/dR$, $f_{RR}\equiv d^2f/dR^2$. Also, configuration space of the (\ref{lag}) Lagrangian is $Q=\{A, B, R\}$. \textbf{In this study we are taking the matter Lagrangian density as $\mathcal {L}_m=-\rho$ for the perfect fluid. Taking the perfect-fluid equation of state of the form $p=w \rho$ the usual conservation equation gives the matter density as $\rho= \rho_{0} A^{-\frac{1+w}{2w}}$ for the metric (\ref{sss-metric})}. As it is known, the equation of state parameter takes the $w=\frac{1}{3}$, $w=0$ and $w=-1$ values respectively for the radiation, matter, and cosmological constant dominate cases. Also, $w=-\frac{1}{3}$ related to a gas of cosmic strings and $w=1$ to stiff matter. Here, due to the restriction from Lagrange, only cases where $w=0$ can be examined.
From the variation with respect to the configuration space elements $A$, $B$ and $R$ for the Lagrangian (\ref{lag}), the field equations of the theory of gravity obtained are as follows:

\begin{eqnarray}\label{alan1}
& &\frac{f_{R}}{4} \, \left(2 \frac{A''}{A} +2\frac{A'B'}{AB}-\frac{A'^2}{A^2}\right)-\left(f_{RR} R'\frac{B'}{B}+f_{RR}R''+f_{RRR}R'^2\right)\nonumber \\
& &-\frac{f}{2}+\frac{1}{w}\kappa \rho_0A^{-\frac{1+w}{2w}}=0,
\end{eqnarray}
\begin{eqnarray}\label{alan2}
& &-\frac{f_{R}}{4} \, \left(2 \frac{A''}{A}-\frac{A'^2}{A^2}+ 4\frac{B''}{B}-2\frac{B'^2}{B^2}\right)+\frac{f_{RR} R'}{2}\left(\frac{A'}{A}+2\frac{B'}{B}\right)\nonumber \\
& &+\frac{f}{2}-\kappa  \rho_0A^{-\frac{1+w}{2w}}=0,
\end{eqnarray}
\begin{eqnarray}\label{alan3}
& &-\frac{f_{R}}{4} \, \left(2 \frac{B''}{B}+\frac{A'B'}{AB}\right)+\frac{f_{RR} R'}{2}\left(\frac{A'}{A}+\frac{B'}{B}\right)\nonumber \\
& &+\frac{f}{2}+f_{RR}R''+f_{RRR}R'^2-\kappa \rho_0A^{-\frac{1+w}{2w}}=0.
\end{eqnarray}

These equations are fourth-order nonlinear differential equations and it is not easy to find an exact solution without making any approach. In this study, we use the Noether symmetry approach given in Appendex \ref{sec:NGS} to find a solution.
First, it is aimed to search for Noether symmetric solutions in GR case for $w=-1$ with $f_0, \rho_0 >0$ condition.
One can obtain the following Noether symmetries by applying the Noether symmetry conditions in the Appendix \ref{SCS1_GR_w-1}.

\begin{eqnarray}\label{s_i}
& &   \textbf{X}_1=\partial_r, \nonumber \\
& &   \textbf{X}_2=e^{\ell r/2}\left[\partial_r+\frac{\ell}{3}\left(A \partial_A+B \partial_B\right)\right],      G=-\frac{\alpha f_0}{3} \sqrt{A}B \ell^2e^{\ell r/2}  ,\nonumber \\
& &   \textbf{X}_3=e^{-\ell r/2}\left[\partial_r-\frac{\ell}{3}\left(A \partial_A+B \partial_B\right)\right],       G=-\frac{\alpha f_0}{3} \sqrt{A}B \ell^2e^{-\ell r/2} ,\nonumber \\
& &   \textbf{X}_4=e^{\ell r/4}\frac{\sqrt{A}}{B^{1/4}} \partial_A, \qquad   \qquad \qquad \qquad    G=-\frac{\alpha f_0}{3} B^{3/4} \ell e^{\ell r/4} ,\nonumber \\
& &  \textbf{X}_5=e^{-\ell r/4}\frac{\sqrt{A}}{B^{1/4}} \partial_A, \qquad  \qquad \qquad  \quad     G=-\frac{\alpha f_0}{3} B^{3/4} \ell e^{-\ell r/4}, \nonumber \\
& &   \textbf{X}_6=e^{\ell r/4}\left(-\frac{A}{2B^{3/4}}\partial_A+B^{1/4} \partial_B\right),    G=-\frac{\alpha f_0}{2} \sqrt{A}B^{1/4} \ell e^{\ell r/4} ,\nonumber \\
& &   \textbf{X}_7=e^{-\ell r/4}\left(-\frac{A}{2B^{3/4}}\partial_A+B^{1/4} \partial_B\right),      G=-\frac{\alpha f_0}{2} \sqrt{A}B^{1/4} \ell e^{-\ell r/4} , \nonumber \\
& &   \textbf{X}_8=2A \partial_A-B\partial_B,
\end{eqnarray}
where $\ell=\sqrt{\frac{6\kappa \rho_0}{f_0}}$. The following Noether integrals corresponding to each Noether symmetry are determined,
\begin{eqnarray}\label{int_i}
& &   I_1=-E_{\mathcal{L}}=0, \nonumber \\
& &   I_2=-e^{\ell r/2}\frac{\ell}{3}\alpha f_0 \sqrt{A}B\left(\frac{A'}{A}+2\frac{B'}{B}-\ell\right),\nonumber \\
& &   I_3=e^{-\ell r/2}\frac{\ell}{3}\alpha f_0 \sqrt{A}B\left(\frac{A'}{A}+2\frac{B'}{B}+\ell\right),\nonumber \\
& &   I_4=-e^{\ell r/4}\alpha f_0 B^{3/4}\left(\frac{B'}{B}-\frac{\ell}{3}\right),\nonumber \\
& &   I_5=e^{-\ell r/4}\alpha f_0 B^{3/4}\left(\frac{B'}{B}+\frac{\ell}{3}\right),\nonumber \\
& &   I_6=-e^{\ell r/4}\alpha f_0 \sqrt{A}B^{1/4}\left(\frac{A'}{A}+\frac{B'}{2 B}-\frac{\ell}{2}\right), \nonumber \\
& &   I_7=-e^{-\ell r/4}\alpha f_0 \sqrt{A}B^{1/4}\left(\frac{A'}{A}+\frac{B'}{2 B}+\frac{\ell}{2}\right), \nonumber \\
& &   I_8=\alpha f_0 \sqrt{A}B\left(\frac{A'}{A}-\frac{B'}{B}\right),
\end{eqnarray}
Using the Noether integrals given above, the following relationships comes out.
\begin{eqnarray}\label{}
& &   I_2e^{-\ell r/2}+I_3e^{\ell r/2}=\frac{2}{3}\ell^2\alpha f_0\sqrt{A}B,\nonumber \\
& &   I_6e^{-\ell r/4}-I_7e^{\ell r/4}=\ell^2\alpha f_0\sqrt{A}B^{1/4},\nonumber \\
& &   \frac{I_4e^{-\ell r/4}-I_5e^{\ell r/4}}{\ell\alpha f_0}=\frac{I_2e^{-\ell r/2}+I_3e^{\ell r/2}}{I_6e^{-\ell r/4}-I_7e^{\ell r/4}},
\end{eqnarray}
Then, $A(r)$ and $B(r)$ metric coefficients are obtained as follows.
\begin{eqnarray}\label{bulg_i}
& & A(r)=A_0\frac{\left(I_6e^{-\ell r/4}-I_7e^{\ell r/4}\right)^2}{\left(I_4e^{-\ell r/4}-I_5e^{\ell r/4}\right)^{2/3}},\nonumber \\
& &   B(r)=B_0\left(I_4e^{-\ell r/4}-I_5e^{\ell r/4}\right)^{4/3},
\end{eqnarray}
where $B_0=(\frac{3}{2\ell\alpha f_0})^{4/3}$ and $A_0=(\frac{2}{3})^{2/3}\frac{1 }{\ell^2(\ell\alpha f_0)^{4/3}}$. Also, according to these solutions, the conditions $\rho_0=\frac{\ell^2f_0}{6\kappa}$ and $I_4 I_7+I_5 I_6=0$ arise.

After the above stage, when $w=-1$ under the $f(R)=f_0R^n$ choosing, solutions in the with matter and non-matter cases are searched as following subsections.

\subsubsection{$f(R)=f_0R^n$,$w=-1$, $f_0,\rho_0 >0$ case:}

Here, we have investigated solutions with the matter for the case of $w=-1$ at higher orders gravity model than GR.
The Noether symmetry equations and the components of the Noether symmetry vectors obtained for this case are given in the Appendix \ref{Rn1}-\ref{Rn1a}.
Then the Noether symmetries are found as follows,
\begin{eqnarray}\label{s_v}
& &   \textbf{X}_1=\partial_r , \qquad
\textbf{X}_2=-2A\partial_A+B\partial_B.
\end{eqnarray}
The corresponding Noether integrals for these Noether symmetries are$I_1=-E_{\mathcal{L}}=0$ and
\begin{eqnarray}\label{int_v}
& &    I_2=\alpha n f_0\sqrt{A}BR^{n-1}\left[\frac{B'}{B}-\frac{A'}{A}\right],
\end{eqnarray}
The following expression is obtained with the help of constant of motion $I_2$.
\begin{equation}\label{sol_v}
A = A_0 B + e^{-\frac{I_2}{2n\alpha f_0} \int{\frac{dr}{\sqrt{A}BR^{(n-1)}}} },
\end{equation}
where $A_0$ is the integration constant for this case.

It is obvious that when $n = 1$, it will be reduced to \textbf{the Einstein theory}. Thus, it is clearly seen that the solution is obtained independently of $R$ as follows.

\begin{equation}\label{sol_v1}
A = A_0 B + e^{-\frac{I_2}{2\alpha f_0} \int{\frac{dr}{\sqrt{A}B}} },
\end{equation}

\subsubsection{$ f(R)=f_0R^n $, $f_0 >0$, $\rho_0 =0$  case:}

Here, we searched for non-matter solutions for the $w=-1$ case in the gravitational model with higher order than GR.
From the Noether symmetry vectors giving in Appendix \ref{Rn1b}, the Noether symmetries are determined as follows;
\begin{eqnarray}\label{sim_iic}
& &   \textbf{X}_1=r\partial_r +2(2n-1)A\partial_A-2R\partial_R, \nonumber\\
& & \textbf{X}_2=\partial_r \qquad \textbf{X}_3=-2A\partial_A+B\partial_B,
\end{eqnarray}

The Noether integrals for the above symmetries are $I_2=-E_{\mathcal{L}}=0$ and

\begin{eqnarray}\label{int_iic}
& & I_1=2\alpha n f_0\sqrt{A}BR^{n-1}\left[-\frac{B'}{B}+(n-1)\frac{A'}{A}-(2n-1)(n-1)\frac{R'}{R}\right], \nonumber\\
& & I_3=\alpha n f_0\sqrt{A}BR^{n-1}\left[\frac{B'}{B}-\frac{A'}{A}\right].
\end{eqnarray}

Thus, (\ref{int_iic}) with the help of the constants of motion, the following expression is obtained.
\begin{equation}\label{sol_iic}
A =A_0 B^\frac{I_1+I_3}{I_1+(n-1)I_3} R^\frac{I_3(2n-1)(n-1))}{I_1+(n-1)I_3} ,
\end{equation}
where $A_0$ is the integration constant for this case.

Here, when \textbf{the $n=1$ condition} is examined, it is seen that the solution is obtained independently from $R$, as in the previous case.

\begin{equation}\label{sol_iic1}
A =A_0 B^{1+\frac{I_3}{I_1} } ,
\end{equation}

\subsection{Static plane symmetric space time}\label{ssec:FEq2}
The plane-symmetrical space-time metric is expressed as follows,
\begin{equation}\label{plsym_metric}
ds^2=A(x)dt^2-B(x)dx^2-Y(x)( dy^2 + dz^2),
\end{equation}
where $A, B$, and $Y$ are functions that depend on $x$.
Then, the following Lagrangian is obtained with the help of the lagrange multiplier and ingration by parts methods.
\begin{eqnarray}\label{plsym_lag}
&&\mathcal{ L}=-2f_{RR}\sqrt{ \frac{A }{B}}R'Y'-f_{RR} \frac{Y }{\sqrt{AB}}A'R' -f_{R}\frac{1}{\sqrt{AB}}A'Y' \nonumber\\
& & \qquad -\frac{f_{R}}{2}\sqrt{\frac{A}{B}}{Y'^2} -\sqrt{AB}Y[Rf_{R}-f+\kappa \rho_{0} A^{-\frac{1+w}{2w}}],
 \nonumber\\
& &\qquad  (w \neq 0)
\end{eqnarray}
According to the conservation law of ${T^{ab}}_{;b}=0$, the energy density expression is determined as $\rho=\rho_{0}A^\frac{-(1+w)}{2w}$. Then, the following field equations are obtained for the theory of gravity expressed by the (\ref{plsym_lag}) Lagrangian.

\begin{eqnarray}
&&\frac{f}{2}B-\frac{f_{R}}{4}\,\left(2\frac{A''}{A}-2\frac{A'Y'}{AY}-\frac{A'^2}{A^2}-\frac{A'B'}{AB}\right)\nonumber\\
& &+\frac{f_{RR} R'}{2}\,\left(2\frac{Y'}{Y}-\frac{B'}{B}\right)+f_{RR}R'' +f_{RRR}R'^2\nonumber\\
& & +\kappa \rho_0BA^{-\frac{1+w}{2w}}=0, \nonumber\\
&&\frac{f}{2}B+\frac{f_{R}}{4}\,\left(-2\frac{A''}{A}+\frac{A'^2}{A^2}-4\frac{Y''}{Y}+\frac{A'B'}{AB}
+2\frac{Y'B'}{YB}\right)\nonumber\\
& &+\frac{f_{RR} R' }{2}\,\left(\frac{A'}{A}+2\frac{Y'}{Y}\right) \nonumber\\
& &+\frac{\kappa }{2}\rho_0BA^{-\frac{1+w}{2w}}=0,\nonumber\\ &&\frac{f}{2}B-\frac{f_{R}}{4}\,\left(2\frac{Y''}{Y}-\frac{B'Y'}{BY}-\frac{A'Y'}{AY}\right)\nonumber\\
& &+\frac{f_{RR} R' }{2}\,\left(\frac{A'}{A}-\frac{B'}{B}+\frac{Y'}{Y}\right)+f_{RR}R''+f_{RRR}R'^2  \nonumber\\
& &+ \kappa \rho_0BA^{-\frac{1+w}{2w}}=0,
\end{eqnarray}
where, taking the variation of the Lagrangian (\ref{plsym_lag}) according to the metric coefficients $A$, $B$ and $Y$ above field equations and variation according to $R$ Ricci scalar expression is obtained. Also, the energy function for the (\ref{plsym_lag}) Lagrangian is obtained in the following form.
\begin{eqnarray}
&&E_\mathcal{L}=\frac{f B}{2} + \frac{f_{R}}{4}\,\left(-\frac{2 A''}{A} + \frac{A'^2}{A^2} - \frac{4 Y''}{Y} + \frac{A'B'}{AB} + \frac{2 Y'B'}{YB}\right)\nonumber\\
& & \qquad+\frac{f'_{R}}{2}\,\left(\frac{A'}{A} + \frac{2 Y'}{Y}\right) +\frac{\kappa \rho_0}{2} B A^{-\frac{1+w}{2w}} ,
\end{eqnarray}
where it is defined as $f'_R=f_{RR}R'$. Since this expression corresponds to the field equation above, it should be $E_\mathcal{L}=0$. In order to find solutions to the equations of this theory, the Noether symmetry approach will be used.

After this stage, to find a solution to the above system of differential equations, Noether symmetries will be examined choosing the function as $ f(R)=f_0R^n $, with the matter case and the cosmological constant dominated non-matter case.

\subsubsection{$ f(R)=f_0R^n $, $f_0, \rho_0 >0$ case}

Here, the search for solutions based on Noether symmetry for the $f(R)=f_0R^n$ model with matter is considered.
Using the Noether symmetric expressions giving Appendix \ref{sec:SPS}-\ref{Rn2}, the Noether symmetries are determined as follows,
\begin{eqnarray}\label{s_plsym}
& & \textbf{X}_1=F(x)\partial_x-2 F_x\partial_B, \nonumber \\
& & \textbf{X}_2=\frac{4wn}{2n-w-1}A\partial_A+\frac{2(w+1)}{2n-w-1}B\partial_B+Y\partial_Y\nonumber\\
& & \qquad -\frac{2(w+1)}{2n-w-1}R\partial_R, \qquad     .
\end{eqnarray}
The corresponding Noether integrals for each Noether symmetry are obtained as follows.
\begin{strip}
\begin{eqnarray}\label{int_plsym}
& &   I_1=-F(x)E_{\mathcal{L}}-2 F_x\frac{\partial_{\mathcal{L}}}{\partial_{B'}}=0, \\
& & I_2=\frac{f_0nR^{n-1}}{2n-w-1}\sqrt{\frac{A}{B}}Y \Big[(2n-w-1)\frac{A'}{A}-2(n-1)(2n-1)(w+1)\frac{R'}{R}+[2n-3(w+1)]\frac{Y'}{Y}\Big],\nonumber
\end{eqnarray}
Using the $I_2$ Noether integral given by (\ref{int_plsym}), the following expression is obtained.
\begin{equation}\label{solplanesymR^n1}
A= A_0 Y^{-\frac{2n -3(w+1)}{(2n-w-1)}}R^{\frac{-2(n-1)(2n-1)(w+1)}{(2n-w-1)}}+e^{\frac{I_2}{nf_0}\int{\sqrt{\frac{B}{A}}\frac{dx}{R^{n-1}Y}}},
 \end{equation}
where $A_0$ is the integration constant for this case.

Once again, it is obvious that in the case of $n=1$ the solution is independent of $R$, and if $w=-1$ is taken, the solution will be obtained in the same order as in the next case.

\begin{equation}\label{solplanesymR^n11}
A=A_0 Y^{-\frac{(3w+1)}{(1-w)}}+e^{\frac{I_2}{f_0}\int{\sqrt{\frac{B}{A}}\frac{dx}{Y}}}
 \end{equation}

\subsubsection{$ f(R)=f_0R^n $, $w=-1$, $f_0, \rho_0 >0$ case:}

 Here, the solutions for $f(R)=f_0R^n$ in cosmological constant dominated vacuum case is investigated. The components of the Noether symmetry vectors are given in Appendex \ref{Rn2w-1rho0}.
Thus, the Noether symmetries are determined as follows,
\begin{eqnarray}\label{s_plsym_Rn}
& & \textbf{X}_1=F(x)\partial_x+2 B F_x \partial_B, \nonumber \\
& & \textbf{X}_2=\frac{2B}{2n-1} \partial_B +Y\partial_Y+\frac{-2R}{2n-1}\partial_R     \nonumber \\
& & \textbf{X}_3=\left( 2(2n-1)A \ln{R} -\frac{2n-3}{n-1}A\ln{Y}+\frac{A}{n-1}\ln{A}\right)\partial_A -Y\left( (2n-1)\ln{R}-\ln{A}+\frac{1}{n-1}\ln{Y}\right)\partial_Y\nonumber \\
& &  \qquad +\frac{B}{n-1}\ln{\frac{A}{Y}}\partial_B-\frac{R}{n-1}\ln{\frac{A}{Y}}\partial_R     \nonumber \\
& &  \textbf{X}_4=A\partial_A +\frac{B}{2n-1}\partial_B -\frac{R}{2n-1}\partial_R.
\end{eqnarray}

Then, for each Noether symmetry above, the corresponding Noether integrals are obtained as follows,
\begin{eqnarray}\label{int_plsym_Rn}
& &   I_1=-E_{\mathcal{L}}=0, \nonumber \\
& &  I_2=-\sqrt{\frac{A}{B}}YnR^{n-1}\left[ \frac{A'}{A}+\frac{Y'}{Y}-\frac{2(n-1)}{2n-1}\left( \frac{A'}{A}+2\frac{Y'}{Y}\right)+(n-1)\frac{R'}{R}\right] ,\nonumber \\
& & I_3=\sqrt{\frac{A}{B}}YnR^{n-1} \Big[\left( \ln{\frac{Y^2A^{\frac{n-1}{n}}}{R^{2n-1}}}\frac{Y'}{Y}+\ln{\frac{Y^{n-1}R^{2n-1}}{A}}\frac{A'}{A}\right)                +(n-1)\ln{\frac{A}{Y}}\left( -(2n-1)\frac{R'}{R}+2\frac{Y'}{Y}+\frac{A'}{A}\right)  \Big] ,\nonumber \\
& &  I_4=-\sqrt{\frac{A}{B}}YnR^{n-1}\left[ \frac{Y'}{Y}+(n-1)\frac{R'}{R}-\frac{(n-1)}{2n-1}\left( \frac{A'}{A}+\frac{Y'}{Y}\right) \right].
\end{eqnarray}
Using the conservative quantities $I_2$ and $I_4$ above, the general solution is determined as follows.
\begin{equation}\label{solplanesymR^n2}
A=A_0 Y^{\frac{-I_2(n-1)+I_4(2n-3)}{I_2(n-1)+I_4}}R^{\frac{(I_2-I_4)(n-1)(2n-1)}{I_2(n-1)+I_4}},
\end{equation}
where $A_0$ is the integration constant for this case.
\end{strip}

In the General Relativity limit where $n = 1$, the solution is as follows.
\begin{equation}\label{solplanesymR^n21}
A=A_0Y^{-1},
\end{equation}

\section{Conclusions}  \label{conc}
In this study, within the scope of $f(R)$ gravity theory, general solutions for static cylindrical and plane symmetric space-times were investigated and the Noether symmetry method was used to obtain exact solutions. As novel research compared to relevant studies in the past, we wanted to determine general solutions instead of compact objects or gravitational wave analysis, which are handled with various approaches.

First, in Sec. I, we have constructed the theory and written the Lagrangian (\ref{lag}) in static cylindrically background. Then, the field equations of the theory have been obtained as (\ref{alan1}), (\ref{alan2}) and (\ref{alan3}).
The Noether symmetry method has been explained in Appendex \ref{sec:NGS}. In this way, the Noether symmetries (\ref{s_i}) and first integrals (\ref{int_i}) have been obtained. Then the solutions for A(r) and B(r) metric coefficient (\ref{bulg_i}) have been obtained in $w=-1$ with $f_0, \rho_0 >0$ condition by using the relation between the first integrals. These solutions are obtained in $f(R) = f_0 R$, namely, in GR case. We wanted to mention this case because this study is an ongoing part of our previous work and for doing a complete analysis. Then, taking $f(R)=f_0R^n$, the solutions for the cosmological constant dominant case $w=-1$ are in the form of (\ref{sol_v}) general expressions.

The next part of this study, is studied within the scope of plane symmetrical space-time. We constructed the theory and written the Lagrangian (\ref{plsym_lag}) in this background. Again, in case of $f(R)=f_0R^n$ selection and $\rho_0 >0$ (\ref{solplanesymR^n1} ) general solutions are determined. Then, the $w=-1$ situation is examined and the general solution expressions (\ref{solplanesymR^n2}) were determined.

In addition, the n=1 GR limits of the solutions are examined and it is seen that they were compatible with the literature.

These obtained solutions point to general expressions obtained with the help of the Noether symmetry method under various conditions. Thus, one can examine these results for various analyses such as compact objects. In this sense, our study is remarkable in terms of its inclusive nature.

\section*{Acknowledgement}
We are grateful to Prof. Salvatore Capozziello for fruitful discussions and valuable comments.
The work of KB was supported in part by the JSPS KAKENHI Grant Number JP21K03547.
\maketitle
\section*{Appendix}
\appendix
\begin{appendix}\label{appendix}
\section{Noether Symmetry Approach}
\label{sec:NGS}
Noether symmetry approach has been widely used in the literature to find exact solutions, particularly of modified gravity theories \cite{Capozziello2007,Capozziello1996,Capozziello2008,Hussain2011,Jamil2011,Kucuakca2011,Vakili2008,Shamir2017,Oz2017} . It opens the way of the solution by decreasing the degrees of freedom of the dynamic system and/or determining the unknown functions of the system. In a sense, the existence of a Noether symmetry is a kind of choice rule. The outline of the Noether symmetry method is given below.
The Lagrangian-related Noether vector $X$ and the first prolongation vector field $X^{[1]}$ can be built as follows:
\begin{equation}\label{NGSvec}
X \mathcal{L}=\xi(\tau,q^k)\frac{\partial\mathcal{L}}{\partial \tau}+\eta^i(\tau,q^k)\frac{\partial\mathcal{L}}{\partial q^i}, \nonumber
\end{equation}
\begin{equation}\label{NGS1vec}
X^{[1]} \mathcal{L}=X \mathcal{L}+\dot{\eta}^k(\tau, q^l, \dot{q}^l)\frac{\partial\mathcal{L}}{\partial \dot{q}^k }. \nonumber
\end{equation}
Where $q^i$ are the generalized coordinates and express the dynamic variables of the extended theory of gravity. Additionally, as $\dot{\eta}^k(\tau, q^l,\dot{q}^l)=D_\tau\eta^k-\dot{q}^kD_\tau\xi$ is defined and $D_\tau=\partial/\partial\tau+\dot{q}^k\partial/\partial q^k$ is the total derivative operator.
If there is a $G(\tau,q ^ k)$ function for any $\mathcal{L}$ Lagrange function and the following Noether symmetry condition is satisfied,
\begin{equation}\label{NSkoş}
X^{[1]} \mathcal{L}+\mathcal{L}(D_\tau\xi)=D_\tau G.\nonumber
\end{equation}
Another important quantity is the first integral as the conservative quantity to be obtained for each Noether symmetry and is expressed as the following:
\begin{equation}\label{FirstInt}
I=-\xi (\dot{q}^i \frac{\partial \mathcal{L}}{\partial \dot{q}^i} -\mathcal{L})+\eta^i\frac{\partial L}{\partial \dot{q}^i}-G.\nonumber
\end{equation}
The first integrals are very useful in that they provide a solution for the theory's system of differential equations.

\section{Noether symmetry properties}
\subsection{Statical cylindrically symmetric space time}\label{sec:SCS}
\subsubsection{GR case for $w=-1$}\label{SCS1_GR_w-1}

Noether symmetry condition (\ref{NSkoş}) gives the following system of differential equations:

\begin{eqnarray}\label{ngs_GR_w-1}
& & \xi_{,A} =0, \qquad \xi_{,B} =0, \qquad \xi_{,R} =0, \nonumber \\
& &  \alpha \,(  f_{R}
\eta^2_{,r} + B f_{RR} \eta^3_{,r})+\sqrt {A}G_{,A}=0,
\nonumber \\
& &  \alpha\,\left[ f_{R}\left(\eta^1_{,r}+{\frac {A\eta^2_{,r}}{B}} \right) +2 \, A f_{RR} \eta^3_{,r} \right]+\sqrt {A} G_{,B} =0,\nonumber \\
& &   \alpha \, f_{RR} \, ( B\eta^1_{,r} +2\,A\eta^2_{,r} ) +\sqrt {A}G_{,R} =0,
\nonumber \\
& &f_{R} \eta^2_{,A} +B f _{RR} \eta^3_{,A} =0,  \nonumber \\
& &    \frac { f_{R} \eta^1  }{2 A}-\frac { f_{R} \eta^2 }{B} +  f _{RR}\eta^3 +2\,\frac {B f _{R}\eta^1_{B}}{A}+2\, f _{R} \eta^2_{B} \nonumber \\
& & +4\,f _{RR} B\eta^3_{B}- f _{R} \xi_{r}=0,
\nonumber \\
& &  -\frac{f_{R} \eta^1 }{2A}+ f _{RR} \eta^3 + f _{R} \eta^1_{,A} + f _{R}\frac {A }
{B}\eta^2_{,A}+2\,A  f _{RR}\eta^3 _{,A} \nonumber \\
& &+ f _{R} \eta^2_{,B} + f _{RR} B \eta^3 _{,B} - f _{R}\xi _{,r} =0,
\nonumber \\
& & -\frac{f_{RR} \eta^1 }{2A}+\frac{f_{RR} \eta^2 }{B}+ f_{RRR} \eta^3 +\frac{f_{R}\eta^2 _{,R} }{B}+ f _{RR}\eta^3 _{,R} \nonumber \\
& &+ f _{RR} \eta^1_{,A}+2\,\frac
{A f _{RR} \eta^2 _{,A}}{B}- f _{RR} \xi_{,r} =0,
\nonumber \\
& &  \frac {  f_{RR} \eta^1 }{A}+2\, f _{RRR} \eta^3+\frac{f_{R} \eta^1}{A}+\frac{f _{R} \eta^2_{,R} }{B}+2\, f _{RR} \eta^3_{,R} \nonumber \\
& &+\frac {B f _{R} \eta^1 _{,B} }{A}+2\, f _{RR} \eta^2 _{,B} -2\, f _{RR} \xi _{,r} =0,
\nonumber \\
& & \left( -f
	+Rf_{R} +\kappa\,\rho_{{0}} {A}^{-\frac {1+w}{2w}} \right)\left[\frac {\alpha\,B}{2\sqrt {A}}\eta^1+\alpha\,\sqrt {A}\eta^2+ \alpha\,\sqrt {A}B\xi_{,r}\right] \nonumber \\
& & - \frac {	\alpha\,B\kappa\,\rho_{{0}} (1+w) }{\sqrt {A}w}{A}^{-\frac {1+w}{2w}} \eta^1    +\alpha\,\sqrt {A}BR f_{RR} \eta^3  +G_{,r} =0, \nonumber \\
& &  B\eta^1 _{,R} +2\,A\eta^2 _{,R} =0.\nonumber
\end{eqnarray}
Solutions have been studied under various conditions for this system of equations. These situations are discussed separately below.

Under the conditions $ f(R)=f_0R $, $ w=-1 $, $ f_0, \rho_0 >0 $ and the definition of $\ell=\sqrt{\frac{6\kappa \rho_0}{f_0}}$ The components of the Noether symmetry vector are obtained as follows,
\begin{eqnarray}\label{ç_GR_w-1}
& &  \xi=c_1+c_2e^{\ell r/2}+c_3e^{-\ell r/2}, \nonumber \\
& &  \eta^1=\frac{A}{3}\ell(c_2 e^{\ell r/2}-c_3 e^{-\ell r/2})+ \frac{\sqrt{A}}{B^{1/4}}(c_4 e^{\ell r/4}+c_5 e^{-\ell r/4})\nonumber \\
& &    \qquad -\frac{A}{2B^{3/4}}(c_6 e^{\ell r/4}+c_7 e^{-\ell r/4})+2 c_8 A,\nonumber \\
& &   \eta^2=\frac{B}{3}\ell(c_2 e^{\ell r/2}-c_3 e^{-\ell r/2})+B^{1/4}( c_6 e^{\ell r/4}+c_7 e^{-\ell r/4})-c_8 B,\nonumber \\
& &   G=-\frac{\alpha f_0}{3} \sqrt{A}B \ell^2(c_2e^{\ell r/2}+c_3e^{-\ell r/2})\nonumber \\
& &-\frac{\alpha f_0}{3} B^{3/4} \ell(c_4e^{\ell r/4}-c_5e^{-\ell r/4})\nonumber \\
& &  \qquad -\frac{\alpha f_0}{2} \sqrt{A}B^{1/4} \ell(c_6e^{\ell r/4}-c_7e^{-\ell r/4})+c_9.\nonumber
\end{eqnarray}

\subsubsection{General case for $f(R)=f_0R^n$, $f_0,\rho_0 >0$}\label{Rn1}
For the case of $f(R)=f_0R^n$, $f_0,\rho_0 >0$ the components of the Noether symmetry vector are obtained as follows,
\begin{eqnarray}\label{ç_iv}
& & \xi=c_1r+c_2, \qquad \eta^1=c_1\frac{4wn}{1+w}A, \qquad \nonumber \\ & &\eta^2=c_1\frac{2n-w-1}{1+w}B\qquad \eta^3= -c_1 2R, \nonumber \\ & &  G=c_3, \qquad \qquad \qquad \qquad(w\neq-1)\nonumber
\end{eqnarray}

\subsubsection{General case for $f(R)=f_0R^n$,$w=-1$, $f_0,\rho_0 >0$}\label{Rn1a}

For the case of $f(R)=f_0R^n$,$w=-1$, $f_0,\rho_0 >0$ the components of the Noether symmetry vector are obtained as follows,
\begin{eqnarray}
& &   \xi=c_1, \qquad \eta^1=-2c_2A, \qquad  \eta^2= c_2B, \qquad G=c_3.\nonumber
\end{eqnarray}

\subsubsection{General case for $ f(R)=f_0R^n $,$w=-1$, $ f_0>0, \rho_0 =0 $}\label{Rn1b}

In the case $ f(R)=f_0R^n $,$ f_0>0, \rho_0 =0 $ the components of the Noether symmetry vector are as follows,
\begin{eqnarray}
& &   \xi=c_1r+c_2, \qquad \eta^1=2(2n-1)c_1A-2c_3A, \nonumber \\ && \eta^2= c_3B, \qquad  \eta^3= -2c_1R, \qquad G=c_4,\nonumber
\end{eqnarray}


\subsection{Statical plane symmetric space time}\label{sec:SPS}
For the (\ref{plsym_lag}) Lagrangian, the following system of equations is obtained by applying the Noether symmetry condition.
\begin{strip}
\begin{eqnarray}
& & \xi_{,A} =0, \qquad \xi_{,B} =0,\qquad \xi_{,Y} =0 , \qquad \xi_{,R} =0,\nonumber \\
& &   f_{R}\eta^3_{,x} +  f_{RR} Y \eta^4_{,x}+\sqrt {A B}G_{,A}=0,\qquad G_{,B} =0\nonumber \\
& &  f_{R}   \left(\eta^1_{,x}+\frac {A} {Y}\eta^3_{,x} \right) +2\, f_{RR} A \eta^4_{,x}  +\sqrt {AB}G_{,Y} =0, \nonumber \\
& &   f_{RR}\left(Y\eta^1_{,x}+2A\eta^3_{,x}\right)+\sqrt{AB}G_{,R}=0,\nonumber\\
& &    f_{R}\eta^3_{,A} +  f_{RR} Y \eta^4_{,A}=0, \nonumber \\
& &  f_{R}\left( \frac{\eta^1}{2A}-\frac{\eta^2}{2B}- \frac{\eta^3}{Y}+\frac{2Y\eta^1_{,Y}}{A}+2\eta^3_{,Y}-\xi_{,x}\right) +f_{RR}\left(\eta^4+4Y\eta^4_{,Y} \right)=0, \nonumber\\
& &  f_{RR}\left(Y\eta^1_{,R} + 2A\eta^3_{,R}\right)=0, \nonumber\\
& &   f_{R}\eta^3_{,B} +  f_{RR} Y \eta^4_{,B}=0, \nonumber \\
& &  f_{R}\left( -\frac{\eta^1}{2A}-\frac{\eta^2}{2B}+\frac{A\eta^3_{,A}}{Y}+\eta^3_{,Y}+\eta^1_{,A}-\xi_{,x}\right)+f_{RR}\left(\eta^4+4Y\eta^4_{,Y} +2A\eta^4_{,A} \right)=0, \nonumber \\
&&f_{R}\frac{\eta^3_{,R}}{Y}+f_{RR}\left(-\frac{\eta^1}{2A}-\frac{\eta^2}{2B}+\frac{\eta^3}{Y}+\frac{2A\eta^3_{,A}}{Y}+\eta^4_{,R}+\eta^1_{,A}-\xi_{,x}\right)+f_{RRR}\eta^4=0, \nonumber \\
& & f_{R}\left( \frac{\eta^1_{,B}}{A}+\frac{\eta^3_{,B}}{Y}\right)+2f_{RR}\eta^4_{,B}=0, \nonumber\\
& &  f_{RR}\left(Y\eta^1_{,B} + 2A\eta^3_{,B}\right)=0, \nonumber\\
&& f_{R}\left(\frac{\eta^1_{,R}}{A}-\frac{\eta^3_{,R}}{Y}\right)+f_{RR}\left(+\frac{\eta^1}{A}-\frac{\eta^2}{B}+\frac{Y\eta^1_{,Y}}{A}+2\eta^4_{,R}+2\eta^3_{,Y}-2\xi_{,x}\right)+2f_{RRR}\eta^4=0, \nonumber \\
&& \left(-f+Rf_{R}+\kappa\rho_0A^{-\frac{1+w}{2w}}\right)\left[\frac{Y}{2}\left(\sqrt{\frac{B}{A}}\eta^1+\sqrt{\frac{A}{B}}\eta^2\right)+\sqrt{A B}(\eta^3+Y\xi_{,x})\right]\nonumber\\
&&+\sqrt{A B} Y (Rf_{RR}\eta^4-\kappa \rho_0\frac{1+w}{w}A^{-\frac{1+w}{2w}}\eta^1)+G_{,x}=0.\nonumber\\
\end{eqnarray}
\end{strip}
\subsubsection{General case for $ f(R)=f_0R^n $, $f_0, \rho_0 >0 $}\label{Rn2}
Under the conditions of $ f(R)=f_0R^n $, $f_0, \rho_0 >0 $, the solution is searched for the system of differential equations and the components of the Noether symmetry vector are obtained as follows,
\begin{eqnarray}
& &  \xi=F(x), \qquad \eta^1=c_2\frac{4wn}{2n-w-1}A \nonumber\\&& \eta^2=-2 F_x+c_2\frac{2(w+1)}{2n-w-1}B, \nonumber \\
& &   \eta^3=c_2Y \qquad  \eta^4=-c_2\frac{2(w+1)}{2n-w-1}R\qquad   G=c_1.\nonumber
\end{eqnarray}

\subsubsection{General case for $f(R)=f_0R^n $,$w=-1$, $ f_0>0 $ $\rho_0=0 $}\label{Rn2w-1rho0}
Using the conditions $f(R)=f_0R^n $, $ f_0>0 $ $\rho_0=0 $, the components of the Noether symmetry vector are found as follows,
\begin{eqnarray}
& & \xi=F(x), \nonumber \\
& & \eta^1=c_3\left( 2(2n-1)A \ln{R} -\frac{2n-3}{n-1}A\ln{Y}+\frac{A}{n-1}\ln{A}\right)\nonumber\\&& \qquad+c_4 A\nonumber \\
& & \eta^2=2 B F_x+c_2\frac{2B}{2n-1}+c_3 \frac{B}{n-1}\ln{\frac{A}{Y}}+c_4\frac{B}{2n-1} \nonumber \\
& &  \eta^3=c_2Y - c_3Y\left( (2n-1)\ln{R}-\ln{A}+\frac{1}{n-1}\ln{Y} \right)\nonumber \\
& & \eta^4=c_2\frac{-2R}{2n-1}-c_3 \frac{R}{n-1}\ln{\frac{A}{Y}}-c_4\frac{R}{2n-1}\nonumber \\
& &  G=c_1.
\end{eqnarray}

\end{appendix}

\bibliographystyle{ieeetr}

\end{document}